# To Switch or Not To Switch: Understanding Social Influence in Recommender Systems

**Haiyi Zhu\*, Bernardo A. Huberman, Yarun Luon**
Social Computing Group, HP Labs
Palo Alto, California, CA
haiyiz@cs.cmu.edu, {bernardo.huberman, yarun.luon}@hp.com

**ABSTRACT**

We designed and ran an experiment to test how often people's choices are reversed by others' recommendations when facing different levels of confirmation and conformity pressures. In our experiment participants were first asked to provide their preferences between pairs of items. They were then asked to make second choices about the same pairs with knowledge of others' preferences. Our results show that others people's opinions significantly sway people's own choices. The influence is stronger when people are required to make their second decision sometime later (22.4%) than immediately (14.1%). Moreover, people are most likely to reverse their choices when facing a moderate number of opposing opinions. Finally, the time people spend making the first decision significantly predicts whether they will reverse their decisions later on, while demographics such as age and gender do not. These results have implications for consumer behavior research as well as online marketing strategies.

**ACM Classification Keywords**
H.5.3 [Information Interfaces and Presentation]: Group and Organization Interfaces – Collaborative computing, Web-based interaction; K.4.4 [Computers and Society]: Electronic Commerce – Distributed commercial transactions.

**General Terms**
Experimentation.

**INTRODUCTION**
Picture yourself shopping online. You already have an idea about what product you are looking for. After navigating through the website you find that particular item, as well as several similar items, and other people's opinions and preferences about them provided by the recommendation system. Will other people' preferences reverse your own?

Notice that in this scenario there are two contradicting psychological processes at play. On one hand, when learning of other's opinions people tend to select those aspects that confirm their own existing ones. A large literature suggests that once one has a position on an issue, one's primary purpose becomes defending or justifying that position [19]. From this point of view, if others' recommendations contradict their own opinion, people will not take this information into account and stick to their own choices. On the other hand, social influence and conformity theory [8] suggest that even when not directly, personally, or publicly chosen as the target of others' disapproval, individuals may still choose to conform to others and reverse their own opinion in order to restore their sense of belonging and self-esteem.

To investigate whether online recommendations can sway peoples' own opinions, we designed an online experiment to test how often people's choices are reversed by others' preferences when facing different levels of confirmation and conformity pressures. We used Rankr [18] as the study platform, which provides a lightweight and efficient way to crowdsource the relative ranking of ideas, photos, or priorities through a series of pairwise comparisons. In our experiment participants were first asked to provide their preferences between pairs of items. Then they were asked to make second choices about the same pairs with the knowledge of others' preferences. To measure the pressure to confirm people's own opinions, we manipulated the time before the participants were asked to make their second decisions. And in order to determine the effects of social pressure, we manipulated the number of opposing opinions that the participants saw when making the second decision. Finally, we tested whether other factors (i.e. age, gender and decision time) affect the tendency to revert.

Our results show that others people's opinions significantly sway people's own choices. The influence is stronger when people are required to make their second decision later (22.4%) rather than immediately (14.1%) after their first decision. Furthermore, people are most likely to reverse their choices when facing a moderate number of opposing opinions. Last but not least, the time people spend making the first decision significantly predicts whether they will reverse their decisions later on, while demographics such as age and gender do not.

## RELATED WORK

### Confirming Existing Opinions
Confirmation of existing opinions is a long-recognized phenomenon [19]. As Francis Bacon stated several centuries ago [2]:

> *"The human understanding when it has once adopted an opinion (either as being received opinion or as being agreeable to itself) draws all things else to support and agree with it. Although there be a greater number and weight of instances to be found on the other side, yet these it either neglects and despises, or else by some distinction sets aside and rejects"*

This phenomenon can be explained by Festinger's dissonance theory: as soon as individuals adopt a position, they favor consistent over inconsistent information in order to avoid dissonance [11].

A great deal of empirical studies supports this idea (see [19] for a review). Many of these studies use a task invented by Wason [26], in which people are asked to find the rule that was used to generate specified triplets of numbers. The experimenter presents a triplet, and the participant hypothesizes the rule that produced it. The participants then test the hypothesis by suggesting additional triplets and being told whether it is consistent with the rule to be discovered. Results show that people typically test hypothesized rules by producing only triplets that are consistent with them, indicating hypothesis-determined information seeking and interpretation. Confirmation of existing opinions also contributes to the phenomenon of belief persistence. Ross and his colleagues showed that once a belief or opinion has been formed, it can be very resistant to change, even after learning that the data on which the beliefs were originally based were fictitious [21].

### Social conformity
In contrast to confirmation theories, social influence experiments have shown that often people change their own opinion to match others' responses. The most famous experiment is Asch's [1] line-judgment conformity experiments. In the series of studies, participants were asked to choose which of a set of three disparate lines matched a standard, either alone or after 1 to 16 confederates had first given a unanimous incorrect answer. Meta-analysis showed that on average 25% of the participants conformed to the incorrect consensus [4]. Moreover, the conformity rate increases with the number of unanimous majority. According to Latané, the relationship between conformity and group size follows a negative accelerating power function [15]. More recently, Cosley and his colleagues [10] conducted a field experiment on a movie rating site. They found that by showing manipulated predictions, users tended to rate movies toward the shown prediction. Researchers have also found that social conformity leads to multiple macro-level phenomenons, such as group consensus [1], inequality and unpredictability in markets [22], unpredicted diffusion of soft technologies [3] and undermined group wisdom [17].

There are informational and normative motivations underlying social conformity, the former based on the desire to form an accurate interpretation of reality and behave correctly, and the latter based on the goal of obtaining social approval from others [8]. However, the two are interrelated and often difficult to disentangle theoretically as well as empirically. Additionally, both goals act in service of a third underlying motive to maintain one's positive self-concept [8].

Both self-confirmation and social conformity are extensive and strong and they appear in many guises. In what follows we consider both processes in order to understand users' reaction to online recommender systems.

### Online recommender systems
Online recommender systems supplement recommendations provided by peers such as friends and coworkers, experts such as movie critics, and industrial media such as *Consumer Reports* by combining personalized recommendations sensitive to people's interests and independently reporting other peoples' opinions and reviews. One popular example of a successful online recommender system is the Amazon product recommender system [16].

### Users' reaction to recommender system
In computer science and HCI, most research in recommendation systems has focused on creating good and effective algorithms (e.g. [5]). There are fewer investigations of the basic psychological processes underlying the interaction of users with recommendations; and none of them addresses both self-confirmation and social conformity. As we mentioned, Cosley and his colleagues [10] studied conformity in movie rating sites and showed that people's rating are significantly influenced by other users' ratings. But they did not consider the effects of self-confirmation or the effects of different levels of social strength. Schwind et al studied how to overcome users' confirmation bias by providing preference-inconsistent recommendations [24]. However, they represented recommendations as search results rather than recommendations from humans, and thus did not investigate the effects of social conformity. Furthermore, their task was more related to logical inference rather than purchase decision making.

In the area of marketing and customer research, studies about the influence of recommendations are typically subsumed under personal influence and word-of-mouth research [23]. Past research has shown that word-of-mouth plays an important role in consumer buying decisions, and the use of internet brought new threats and opportunities for marketing [23,13,25]. There were several studies specifically investigating the social conformity in product evaluations [7,9,20]. Although they found substantial effects of others' evaluations on people's own judgments, the effects were not always significantly stronger when others' opinions were more uniform[1]. In Burnkrant and Cousineau's [7] and Cohen and Golden's [9] experiments, subjects were exposed to evaluations of coffee with high uniformity or low uniformity. Both results showed that participants did not exhibit significantly increased adherence to others' evaluation in the high uniformity condition (although in Burnkrant's experiments, the participants significantly recognized the difference between high and low uniformity). On the other hand, in Pincus and Waters's experiments (college students rated the quality of one paper plate while exposed to simulated quality evaluations of other raters), it was found that conformity effects are stronger when the evaluations were more uniform[20].

In summary, while previous research showed that others' opinions can influence people's own decisions, none of that research addresses both the self-confirmation and social conformity mechanism that underlie choice among several recommendations. Additionally, although previous researchers concluded that people are more likely to be influenced when the social conformity pressures are stronger (i.e. more people uniformly oppose them), the empirical results were mixed. By contrast our experiments address how often people reverse their own opinions when confronted with others people preferences, especially when facing different levels of confirmation and conformity pressures.

**EXPERIMENTAL DESIGN**
We conducted a series of online between-subjects experiments. All participants were asked to go to the website of Rankr [2] [18] to make a series of pairwise comparisons with or without knowledge of others people's preferences (Figure 1)[3]. We wanted to determine whether people reverse their choices by seeing others' preferences.

---

[1] Unlike other conformity experiments such as line-judgment where the strength of social conformity is manipulated by increasing the number of "unanimous" majority, experiments about social influence in product evaluation [7, 9, 20] usually manipulate the strength of social influence by changing the degree of uniformity of opinions. As discussed in [9], since it is seldom that no variation exists in the advice or opinions in reality, the latter method is more likely to stimulate participants' real reactions. We also use the latter method in our experiment by manipulating the ratio of opposing opinions versus supporting opinions.

[2] http://www.hpl.hp.com/research/scl/papers/rankr/rankr.pdf

[3] The pictures were collected from Google Images.

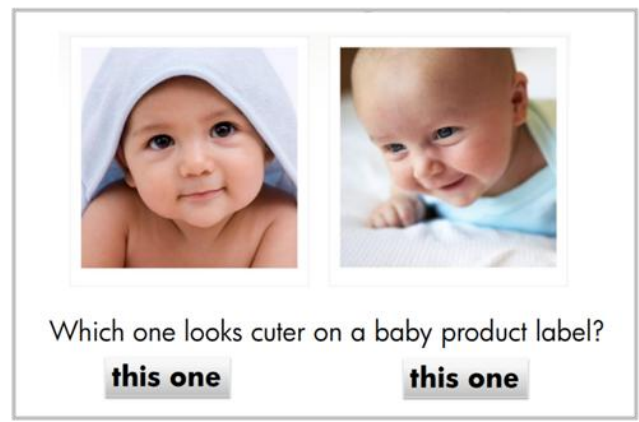

a. Comparing baby pictures, not showing others' preferences

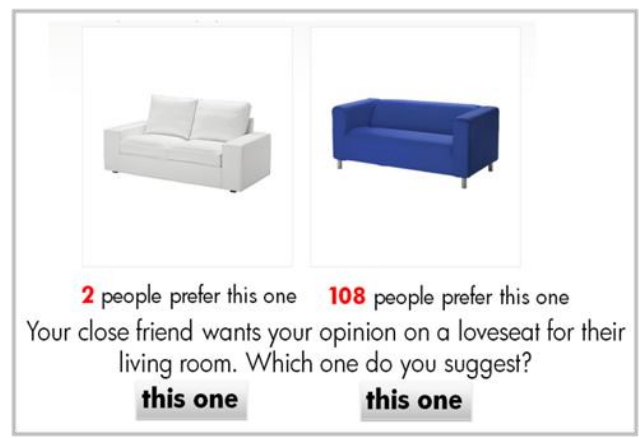

b. Comparing loveseats, showing others' preferences
Figure 1. Example pairwise comparisons in Rankr

**Conditions**
The experimental design was 2x3x4 measuring (baby pictures and loveseat pictures) versus (strong confirmation, weak confirmation and control) versus (ratio of opposing opinions versus supporting opinions: 2:1, 5:1, 10:1 and 20:1). Participants were recruited from Amazon's Mechanical Turk and were randomly assigned into one of six conditions (baby–strong, baby-weak, baby–control, loveseat-strong, loveseat-weak and loveseat-control) and made four choices with different levels of conformity pressure.

In the baby condition, people were asked to compare twenty-three or twenty-four pairs of baby pictures by answering the question "which baby looks cuter on a baby product label". In the loveseat condition, the question was "your close friend wants your opinion on a loveseat for their living room, which one do you suggest"; and people also needed to make twenty-three or twenty-four choices.

In the strong confirmation condition, people first compared two pictures on their own and they were then immediately asked to make another choice with available information about others' preferences. When the memories were fresh, reversion leads to strong inconsistency and dissonance of people choices with their own previous ones. Furthermore, we tested whether people would reverse their first choice under four levels of social pressure: when the number of opposing opinions were twice, five times, ten times, and twenty times as many as the number of people who supported their opinions. The numbers were randomly generated[4]. Except for theses eight experimental pairs, we also added fourteen noise pairs and an honesty test composed of two pairs (twenty-four pairs in total, see Figure 2 for an example). In this condition, noise pairs also consisted of consecutive pairs (a pair with social information immediately after the pair without social information). However, others' opinions were either indifferent or in favor of the participants' choices. We created an honesty test to identify participants who cheated the system and quickly clicked on the same answers. The test consisted of two consecutive pairs with the same items but with the positions of the items exchanged. Participants needed to make the same choices among these consecutive two pairs in order to pass the honesty test. The relative orders of experimental pairs, noise pairs, and honesty test in the sequence and the items in each pair were randomly assigned to each participant.

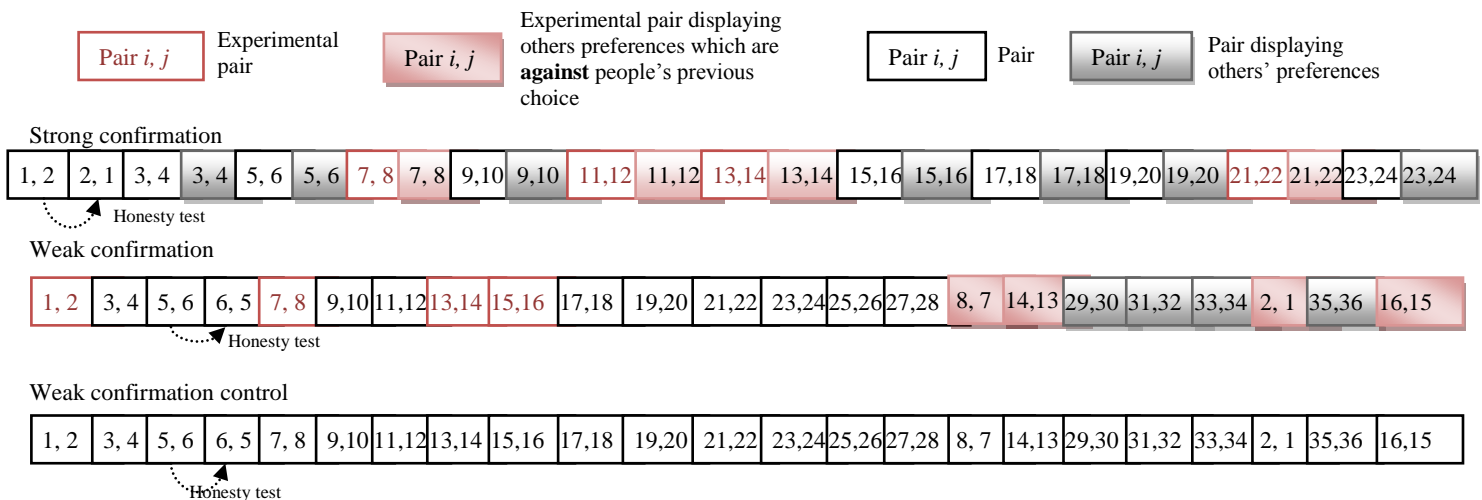

**Figure 2. Example displaying orders in each condition.**

In contrast with the strong confirmation condition where people were aware that they reversed their choices, in the weak confirmation condition we manipulated the order of display and the item positions so that the reversion was less explicit. People first compared pairs of the items without knowing others' preferences, and then on average after 11.5 pairs later we showed the participants the same pair (with the positions of items in the pair exchanged) and others' opinions. Similarly, with the conscious condition we showed eight experimental pairs to determine whether people reversed their previous choices with increasing strength of social influence. Additionally, we showed thirteen noise pairs (nine without others' preferences and four with others' preferences) and performed an honesty test (see Figure 2 for an example).

By manipulating the time between two choices, we blurred people's memories of their choices in order to exert a subtle confirmation pressure. However, as people proceeded with the experiment they were presented with new information to process. This new information may lead them to think in a different direction and change their own opinions regardless of social influence. In order to control for this confounding factor, we added a weak confirmation control condition, where the order of the pairs were the same as with the weak confirmation condition but without showing the influence of others.

**Procedures**

We conducted our experiment on Amazon Mechanical Turk (mTurk) [14]. The recruiting messages stated that the objective of the survey was to do a survey to collect people's opinions. Once mTurk users accepted the task they were asked to click the link to Rankr, which randomly directed them to one of the six conditions. This process was invisible to them.

First, the participants were asked to provide their preferences about twenty-three or twenty-four pairs of babies or loveseats. They were then directed to a simple survey. They were asked to report their age, gender and answer two 5-Likert scale questions[5]. After filling out the survey, a unique confirmation code was generated and displayed on the webpage. Participants needed to paste the code back to the mTurk task. With the confirmation code in hand we matched mTurk users with the participants of our experiments, allowing us to pay mTurk users according to their behaviors. We paid $0.35 for each valid response.

---

[4] We first generated a random integer from 150 to 200 as total participants. Then we generate the number of people holding different opinions according to the ratio. Here are a few examples: 51 vs 103 (2X), 31 vs156 (5X), 16 vs 161 (10X) and 9 vs 181(20X).

**Participants**

We collected 600 responses. Of this number, we omitted 37 responses from 12 people who completed the experiment multiple times; 22 incomplete responses; 1 response which did not conform to the participation requirements (i.e. being at least 18 years old); and 107 responses who did not pass the honesty test. These procedures left 433 valid participants in the sample, about 72% of the original number. According to participant self-reporting, 40% were females; age ranged between18 to 82 with a median age of 27 years. Geocoding[6] the ip addresses of the participants revealed 57% were from India, 25% from USA, with the remaining 18% of participants coming from over 34 different countries.

The numbers of participants in each condition were as follows. Baby-strong: 72; baby-weak: 91; baby-control: 49; loveseat-strong:75; loveseat-weak:99; loveseat-control:47.[7]

People spent a reasonable amount of time on each decision (average 6.6 seconds; median 4.25 seconds).

Among the 433 responses, 243 left comments in the open-ended comments section at the end of the experiments. Most of them said that they had a good experience when participating in the survey. (They were typically not aware that they were in an experiment).

**Measures**

- Reversion: whether people reverse their preferences after knowing others' opinion.

- Strength of social influence: the ratio of opposing opinions to supporting opinions.
- Decision time: the time (in seconds) people spent in making each decision.
- Demographic information: age and gender.
- Self-reported usefulness of others' opinions.
- Self-reported level of being influenced.

**RESULTS**

**Did people reverse their opinions by others' preferences when facing different confirmation pressure?**

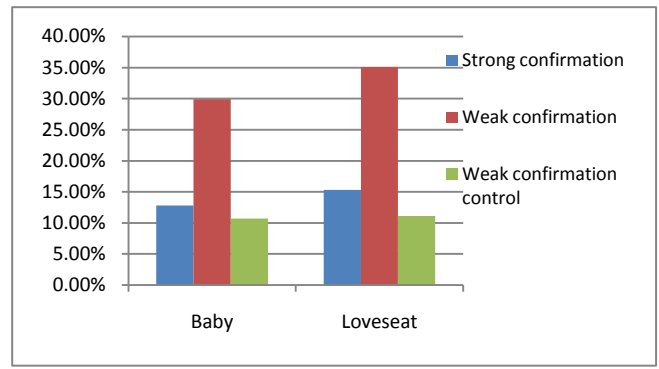

**Figure 3.Reversion rate by conditions.**

Figure 3 shows the reversion rate as a function of the conditions we manipulated in our experiment. First, we found out that content does not matter, i.e., although baby pictures are more emotionally engaging than loveseat pictures, the patterns are the same. The statistics test also shows that there is no significant difference between the baby and the loveseat results ($t(431)=1.35$, $p=0.18$).

Second, in the strong confirmation condition, the reversion rate was 14.1%, which is significantly higher than zero ($t(146)=6.7$, $p<0.001$).

Third, the percentage of people that reversed their opinion was as high as 32.5% in the weak confirmation condition, significantly higher than the weak confirmation control condition (10.1%). This difference is significant: $t(284)=6.5$, $p<0.001$. We can therefore conclude that social influence contributes approximately to 22.4% of the reversion of opinions observed.

To summarize the results, in both the strong and the weak confirmation conditions, others' opinions significantly swayed people's own choices (22.4% and 14.1%). The effect size of social influence was larger when the self-confirmation pressure was weaker.

In order to calibrate the magnitude of our results, we point out that they are of the same magnitude as the classic line-judgment experiments. According to a 1996 meta-analysis

---

[5] The questions were as follows. "Is showing others' preferences useful to you?" "How much does showing others' preferences influences your response?"

[6] MaxMind GeoLite was used to geocode the ip addresses which self-reports a 99.5% accuracy rate.

[7] Among the 600 responses, originally 20% were assigned for baby-strong; 20% for baby-weak; 10% for baby-control; 20% for loveseat-strong; 20% for loveseat-weak and10% for loveseat-control. The valid responses in strong confirmation conditions were fewer than the ones in weak confirmation conditions because the strong confirmation condition had a higher failure rate in the honesty test. The reason might be that strong confirmation condition had more repetitive pairs, fewer new items and more straightforward patterns, leading to boredom and casual decisions, which in turn caused failure in the honesty tests.

of line-judgment experiment consisting of 133 separate experiments and 4,627 participants, the average conformity rate is 25% [4]. Thus the magnitude of our results is consistent with other experimental work on social influence

**Were people more likely to reverse their own preferences when more people are against them?**

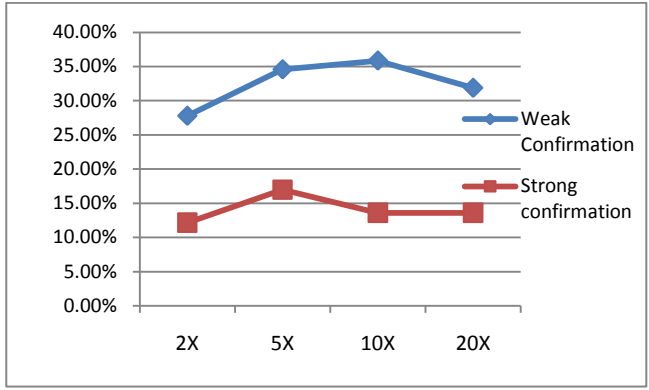

**Figure 3. Reversion rate by the strength of social influence.**

Interestingly, we saw an increasing and then decreasing trend when the opposing opinions became stronger, i.e., the condition with the most uniform opposing opinions (20X) was not more effective in reversing people's own opinions than the moderate opposing opinions (5X and 10X).

These results might be explained by Brehn's finding of psychological reactance [6]. According to Brehn, if an individual's freedom is perceived as being reduced or threatened with reduction, he will become aroused to maintain or enhance his freedom. The motivational state of arousal to reestablish or enhance his freedom is termed psychological reactance. Therefore if the participants perceived the uniform opposing opinions as a threat to their freedom to express their own opinions, their psychological reactance might be aroused to defend and confirm their own opinions.

These results can also be explained in terms of Wu and Huberman's findings about online opinion formation [27]. In their work they used the idea of maximizing the impact that individuals have on the average rating of items to explain the phenomenon that later reviews tend to show a big difference with earlier reviews in Amazon.com and IMDB.

We can use the same idea to explain our results. Social influence in product recommendations is not just a one-way process. People are not just passively influenced by others' opinions but also want to maximize their impact on other people's future decision making (e.g., in our experiments, according to our recruiting messages, participants would assume that their choices would be recorded in the database and shown to others; in real life, people like to influence their friends and family). We assume that the influence of an individual on others can be measured by how much his or her expression will change the average opinion. Suppose there are $X_1$ supporting opinions and $X_2$ opposing opinions, and that $X_2 > X_1$. A person's choice $c$ (0 indicates confirming his or her own choice, 1 indicates conforming to others) can move the average percentage of opposing opinions from $X_2/(X_1 + X_2)$ to $(X_2 + c)/(X_1 + X_2 + 1)$. So the influence on the average opinion is $\left| \frac{X_2}{X_1+X_2} - \frac{X_2+c}{X_1+X_2+1} \right|$. A simple derivation shows that to maximize the influence on average opinion, people need to stick to their own choices and vote for the minority. Then their influence gain will be stronger when the difference between existing majority opinions and minority ones is larger. Therefore, the motivation to exert influence on other people can play a role in resisting the social conformity pressure and lead people to confirm their own decisions especially when facing uniform opposing opinions.

**What else predicts the reversion?**

| Predictors | Coef. | Std. Err. | P>|z| |
|---|---|---|---|
| Condition (1-strong confirmation; 0-weak confirmation) | 1.26 | .152 | <.001 |
| Age | -5.71e-3 | 6.89e-3 | 0.407 |
| Gender | 6.67e-2 | .143 | 0.642 |
| Self-reported usefulness | .164 | .070 | 0.02 |
| Self-reported influence level | .334 | .072 | <.001 |
| Std. first decision time | .323 | .065 | <.001 |
| Log likelihood | | -657.83 | |

**Table 1. Logistic regression predicting the reversion.**

We used a logistic regression model to predict reversion with the participants' age, gender, self-reported usefulness of recommendation system, self-reported level of being influenced by the recommendation systems and standardized first decision time.

The results showed that age and gender do not significantly predict reversion (p=0.407, p = 0.642). Self-reported influence level has a strong prediction power (Coef. = 0.334, p <0.001), which is reasonable. The interesting fact is that decision time, a simple behavioral measure, predicts reversion almost as well as self-reported influence level (Coef. = 0.323, p<0.001). The longer people spent on the decisions, the more equivalent the two choices are for them. According to Festinger's theory [12]: the more equivocal the evidence, the more people rely on social cues. Therefore, the more time people spend on a choice, the more likely they are to reverse this choice and conform to others later on.

**LIMITATIONS & FUTURE WORK**

In our experiments, we examined whether people reverse their choices when facing different ratios of opposing opinions versus supporting opinions (2X, 5X, 10X and 20X). In order to further investigate the relationship between the ratio of the opposing opinions and the tendency to revert, it would be better to include more fine-grained conditions in the ratio of opposing opinions. The ideal situation would be a graph with the continuous opposing versus supporting ratio as the x-axis and the reversion rate as the y-axis.

Also, additional manipulation checks or modification of the design of the experiment would be needed to establish whether processes such as psychological reactance or the intent to influence others have been operating. For example, the degree of perceived freedom in the task could be measured. And it would be revealing to manipulate whether or not people's choices would be visible to other participants to see whether the intention of influencing others takes effect.

In order to collect earnest responses, we used several methods such as honest test and IP address checking. The average time they spent on the task, the statistically significant results of the experiment and the comments participants left all indicate that our results are believable. However, there is still a limitation of our honesty test. On one hand, the honesty test (the consecutive two pairs with positions of items switched) was able to identify all the users who tried to cheat the system by randomly clicking on the results, which added noise in our data. On the other hand, the honesty test might also exclude some earnest responses. It is possible that, immediately after people made a choice, they regret it.

During our research, we invented an experimental paradigm to easily measure and manipulate the number of conditions under which people make choices under some kind of social influence. This paradigm can be extended to scenarios beyond those of binary choices, to the effect of recommendations from friends as opposed to strangers and whether social influence varies with different modalities of recommendation visualizations.

**CONCLUSION**

In this paper, we present results of a series of online experiments designed to investigate whether online recommendations can sway peoples' own opinions. These experiments exposed participants making choices to different levels of confirmation and conformity pressures.

Our results show that people's own choices are significantly swayed by the perceived opinions of others. The influence is weaker when people have just made their own choices. Additionally, we showed that people are most likely to reverse their choices when facing a moderate, as opposed to large, number of opposing opinions. And last but not least, the time people spend making the first decision significantly predicts whether they will reverse their own later on.

Our results have three implications for consumer behavior research as well as online marketing strategies. 1) The temporal presentation of the recommendation is important; it will be more effective if the recommendation is provided not immediately after the consumer has made a similar decision. 2) The fact that people can reverse their choices when presented with a moderate countervailing opinion suggests that rather than overwhelming consumers with strident messages about an alternative product or service, a more gentle reporting of a few people having chosen that product or service can be more persuasive than stating that thousands have chosen it. 3) Equally important is the fact that a simple monitoring of the time spent on a choice is a good indicator of whether or not that choice can be reversed through social influence. There is enough information in most websites to capture these decision times and act accordingly.

**REFERENCES**
1. Asch, S.E. (1956) Studies of independence and conformity: I. A minority of one against a unanimous majority. Psychological Monographs, Vol 70(9), 1956, 70.
2. Bacon, F., (1939), Novum organum. In Burtt, E.A. (Ed), The English philosophers from Bacon to Mill (pp.24-123). New York: Random House. (Original work published in 1620).
3. Bendor. J., Huberman, B.A., Wu,F., (2009) Management fads, pedagogies, and other soft technologies, Journal of Economic Behavior & Organization, Volume 72, Issue 1, October 2009, Pages 290-304
4. Bond, R, Smith, P. B., Culture and conformity: A meta-analysis of studies using Asch's (1952b, 1956) line judgment task. Psychological Bulletin, Vol 119(1), Jan 1996, 111-137.
5. Breese JS, Heckerman D, Kadie C. (1998) Empirical analysis of predictive algorithms for collaborative filtering. Proceedings of the 14th Conference on Uncertainty in Artificial Intelligence. Morgan Kaufmann: San Francisco, CA.
6. Brehm, J.W. (1966) A Theory of Psychological Reactance. New York: Academic Press.
7. Burnkrant, R.E., and Cousineau, A., Informational and Normative Social Influence in Buyer Behavior, Journal of Consumer Research, Vol. 2, No. 3 (Dec., 1975), pp. 206-215.
8. Cialdini, R.B., and Goldstein,N.J., Social Influence: Compliance and Conformity. Annual Review of Psychology. Vol. 55: 591-621.
9. Cohen, Joel B.; Golden, Ellen (1972) Informational social influence and product evaluation. Journal of Applied Psychology, Vol 56(1), Feb 1972, 54-59.

10. Cosley. D., Lam, S.K, Albert., I, Konstan., J.A., and Riedl. J. 2003. Is seeing believing?: how recommender system interfaces affect users' opinions. In Proceedings of the SIGCHI conference on Human factors in computing systems (CHI '03).
11. Festinger, L. (1957). A theory of cognitive dissonance. Stanford University Press.
12. Festinger, L., (1954). A theory of social comparison processes. Human Relations 7, 117–140.
13. Hennig-Thurau, T., Gwinner, K. P., Walsh, G., Gremler, D. D., (2004) Electronic word-of-mouth via consumer-opinion platforms: What motivates consumers to articulate themselves on the Internet? Journal of Interactive Marketing.
14. Kittur, A., Chi, E.H., and Suh, B. (2008) Crowdsourcing user studies with Mechanical Turk. In Proceeding of the twenty-sixth annual SIGCHI conference on Human factors in computing systems (CHI '08). ACM, New York, NY, USA, 453-456.
15. Latané, B.,(1981) The psychology of social impact. American Psychologist, Vol 36(4), Apr 1981, 343-356.
16. Linden, G.; Smith, B.; York, J.; , Amazon.com recommendations: item-to-item collaborative filtering, Internet Computing, IEEE , vol.7, no.1, pp. 76- 80, Jan/Feb 2003
17. Lorenz. J., Rauhut, H., Schweitzer, F., and Helbing, D. How social influence can undermine the wisdom of crowd effect. PNAS 2011 108 (22) 9020-9025.
18. Luon, Y., Aperjis, C., Huberman, B.A., Rankr: A Mobile System for Crowdsourcing Opinions. http://www.hpl.hp.com/research/scl/papers/rankr/rankr.pdf
19. Nickerson, R. S. Confirmation bias: A ubiquitous phenomenon in many guises. Review of General Psychology, Vol 2(2), Jun 1998, 175-220.
20. Pincus, S.; Waters, L. (1977) K. Informational social influence and product quality judgments. Journal of Applied Psychology, Vol 62(5), Oct 1977, 615-619.
21. Ross, L., Lepper, M. R., & Hubbard, M. (1975). Perserverance in self perception and social perception: Biased attributional processes in the debriefing paradigm. Journal of Personality and Social Psychology, 32, 880-892.
22. Salganik, M.J., Dodds, P.S., and Watts. D.J., Experimental Study of Inequality and Unpredictability in an Artificial Cultural Market. Science 10 February 2006: Vol. 311 no. 5762 pp. 854-856.
23. Senecal, S., Nantel, J., The influence of online product recommendations on consumers' online choices, Journal of Retailing, Volume 80, Issue 2, 2004, Pages 159-169.
24. Schwind, C., Buder, J., and Hesse, F.W. 2011. I will do it, but i don't like it: user reactions to preference-inconsistent recommendations. In Proceedings of the 2011 annual conference on Human factors in computing systems (CHI '11).
25. Stauss, B. (1997). Global Word of Mouth. Service Bashing on the Internet is a Thorny Issue. Marketing Management, 6(3), 28 –30.
26. Wason, P. C. (1960) On the failure to eliminate hypotheses in a conceptual task.The Quarterly Journal of Experimental Psychology, Vol 12, 1960, 129-140.
27. Wu, F., and Huberman, B.A., (2010) Opinion formation under costly expression. ACM Trans. Intell. Syst. Technol. 1, 1, Article 5 (October 2010), 13 pages.